\documentclass[api,pra,amsmath,amssymb,twocolumn]{revtex4-1}

\usepackage{bm}
\usepackage[bookmarks=false]{hyperref}

\def\dd{\mathrm{d}}

\def\ii{\mathrm{i}}

\def\muz{\mu_0}

\def\cavlenz{l_0}

\def\oHH{\hat{\mathcal{H}}}

\def\oH{\hat{H}}

\def\oa{\hat{a}}
\def\oad{\hat{a}^{\dagger}}

\def\oalpha{\hat{\alpha}}
\def\oalphad{\hat{\alpha}^{\dagger}}

\begin{document}


\title{Reply to Comment on ``System-environment coupling derived by Maxwell's boundary conditions from the weak to the ultrastrong light-matter-coupling regime''}

\author{Motoaki Bamba}
\altaffiliation{E-mail: bamba@acty.phys.sci.osaka-u.ac.jp}
\affiliation{Department of Physics, Osaka University, 1-1 Machikaneyama, Toyonaka, Osaka 560-0043, Japan}
\author{Tetsuo Ogawa}
\affiliation{Department of Physics, Osaka University, 1-1 Machikaneyama, Toyonaka, Osaka 560-0043, Japan}
\date{\today}

\date{\today}

\begin{abstract}
As mentioned by Simone De Liberato \cite{DeLiberato2013a2},
when we suppose the metallic thin mirror
and perform the renormalization additionally to the approach
starting from the frequently-used system-environment coupling Hamiltonian,
we can certainly resolve the discrepancy of its result from that
obtained by the reliable approach in the main discussion of our paper \cite{Bamba2013MBC}.
Although the suggested approach is currently applicable to the specific situation
after checking its validity
by our reliable but cumbersome approach,
we instead propose to start from the system-environment coupling Hamiltonian
determined properly by the mechanism of the confinement and loss of the cavity fields.
This approach is applicable to any cavity structures in principle,
and we do not face the renormalization problem appearing in the comment.
\end{abstract}


\maketitle
In our paper \cite{Bamba2013MBC},
we calculated the loss rates of cavity polaritons
by the Maxwell's boundary conditions (reliable approach)
and by the approach starting from the Hamiltonian of system-environment coupling
\begin{equation} \label{eq:HSE_standard} 
\oHH_{\text{S-E}}^{\text{standard}} = \sum_m\int\dd\omega\
\ii\hbar\sqrt{\frac{\kappa_m(\omega)}{2\pi}}
\left[ \oalphad(\omega)\oa_m - \oad_m\oalpha(\omega) \right],
\end{equation}
which usually supposed
in the study of cavity quantum electrodynamics (QED).
We found that,
in the ultrastrong light-matter coupling regime, the latter approach
[called ``the rotating-wave approximation (RWA) approach'' in the comment]
gives inconsistent loss rates compared with those calculated
by the reliable approach based on the Maxwell's boundary conditions.
Then, we concluded that Eq.~\eqref{eq:HSE_standard}
is unreliable in the ultrastrong light-matter coupling regime.

In the comment by Simone De Liberato \cite{DeLiberato2013a2},
he suggests that there are two points to be reconsidered
in the demonstrations of our paper.
Point (I): In the RWA approach,
the loss rates of the eigen-modes (lower and upper polaritons)
should be properly renormalized with considering the normalization condition
of the Hopfield-Bogoliubov coefficients $\{w_j, x_j, y_j, z_j\}$.
Point (II): The permittivity of the thin mirror should be proportional to
$\eta(\omega) \propto \omega^{-2}$ (Drude model, i.e., the mirror should be metallic),
whereas we supposed $\eta(\omega) \propto \omega^{-1}$
just for simplicity without considering any physical justification.
Taking into account these two points,
the RWA approach starting from Eq.~\eqref{eq:HSE_standard}
certainly reproduces the loss rates calculated reliably
from the Maxwell's boundary conditions
in the main discussion of our paper \cite{Bamba2013MBC}.
Briefly speaking, the comment \cite{DeLiberato2013a2} shows
that, in certain situations, Eq.~\eqref{eq:HSE_standard} is still usable 
even in the ultrastrong light-matter coupling regime.

Let us first discuss Point (II). 
Since we supposed the infinitely thin mirror
[not distributed Bragg reflectors (DBRs)],
as pointed out in the comment,
it is natural to suppose that the mirror is made of metal
when we consider the actual experimental realizations
of the ultrastrong light-matter coupling
by the intersubband transitions in semiconductor quantum wells \cite{Gunter2009N}
and by the molecular materials \cite{Schwartz2011PRL}.
Then, it is certainly reasonable to suppose $\eta(\omega) \propto \omega^{-2}$,
if the permittivity of the thin mirror obeys the Drude model
in the wide frequency range of interest.
For other cavity structures, e.g.,
thin mirrors not obeying the Drude model, DBRs,
subwavelength structures \cite{Todorov2009PRL},
split-ring resonators \cite{Scalari2012S},
and transmission line resonators in superconducting circuits
\cite{Johansson2010PRA},
we should reconsider the fundamental mechanisms
of the confinement and loss of the cavity fields \cite{Bamba2013SEC}.

Concerning Point (I),
in our paper \cite{Bamba2013MBC},
we did not perform the renormalization suggested by the comment,
and then the RWA approach starting from Eq.~\eqref{eq:HSE_standard}
gave inconsistent results with the reliable ones.
This is the reason why we concluded this approach unreliable.
But,
why do we need additionally such renormalization
of the loss rates calculated from Eq.~\eqref{eq:HSE_standard}
to reproduce those calculated by the Maxwell's boundary conditions?
As seen in Eq.~(5) of the comment,
the loss rates of polaritons are renormalized
to keep them lower than the bare loss rate $\kappa_m(\omega)$ of cavity photons.
The origin of this additional treatment seems
the starting point of the RWA approach,
i.e., Eq.~\eqref{eq:HSE_standard}.
This Hamiltonian was supposed in our paper \cite{Bamba2013MBC}
and also in the comment \cite{DeLiberato2013a2}, and
it is obtained by performing the RWA
on the ``original'' system-environment coupling 
Hamiltonian in the basis of ``photons'' not of ``polaritons''.
In our recent work \cite{Bamba2013SEC},
we derived the system-environment coupling Hamiltonian
based on the fundamental mechanism of the confinement and loss
of the cavity fields.
The ``original'' Hamiltonian for the cavity structure supposed in our paper \cite{Bamba2013MBC}
is derived as \cite{Bamba2013SEC}
\begin{equation} \label{eq:HSE_H_FP} 
\oHH_{\text{S-E}} = \int\dd\omega\
\ii\hbar\sqrt{\frac{\muz c^3}{\pi\hbar\omega^3\eta(\omega)^2}}
\oH(z=0^+)\left[ \oalphad(\omega) - \oalpha(\omega) \right],
\end{equation}
where $\oH(z=0^+)$ is the operator of magnetic field inside the cavity
at the cavity mirror ($z = 0$).
When the ultrastrong light-matter coupling is mediated by the electric dipoles,
Eq.~\eqref{eq:HSE_H_FP} can be rewritten
in the Coulomb gauge and under the definition of $\oa$ in our paper \cite{Bamba2013MBC} as
\begin{subequations} \label{eq:HSE_a_FP} 
\begin{align}&
\oHH_{\text{S-E}}
\nonumber \\ 
& = \sum_m\int\dd\omega\
\ii\hbar\sqrt{\frac{\kappa_0(\omega)}{2\pi}\frac{\omega_m}{\omega}}
(\oa+\oad)\left[ \oalphad(\omega) - \oalpha(\omega) \right], \\
& = \sum_m\int\dd\omega\
\hbar\sqrt{\frac{\kappa_0(\omega)}{2\pi}\left(\frac{\omega_m}{\omega}\right)^3}
(\oa-\oad)\left[ \oalphad(\omega) - \oalpha(\omega) \right],
\end{align}
\end{subequations}
where the bare loss rate is defined as
\begin{equation}
\kappa_0(\omega) = \frac{2c^3}{\omega^2\eta(\omega)^2\cavlenz}.
\end{equation}
According to ``the RWA approach'' starting from Eqs.~\eqref{eq:HSE_a_FP}
not from Eq.~\eqref{eq:HSE_standard},
the loss rates of lower ($j = L$) and upper polaritons ($j = U$) are represented as
\begin{subequations}
\begin{align}
\kappa_j
& = \kappa_0(\omega_j) \left(\frac{\omega_m}{\omega_j}\right) |w_j-y_j|^2 \\
& = \kappa_0(\omega_j) \left(\frac{\omega_m}{\omega_j}\right)^3 |w_j+y_j|^2.
\end{align}
\end{subequations}
These expressions certainly reproduces the loss rates
calculated by the Maxwell's boundary conditions \cite{Bamba2013MBC}
for arbitrary model of permittivity $\eta(\omega)$ of the thin mirror.
Furthermore, starting from Eqs.~\eqref{eq:HSE_a_FP} [or more generally Eq.~\eqref{eq:HSE_H_FP}],
we do not need the renormalization that is performed in the comment.
This fact implies that the renormalization problem discussed in the comment
appears because the starting point Eq.~\eqref{eq:HSE_standard} is obtained
by applying the RWA on Eqs.~\eqref{eq:HSE_a_FP} in the photon basis
and then the RWA is applied again on Eq.~\eqref{eq:HSE_standard}
in the polariton basis to calculate the loss rates.
When we apply the RWA directly on the ``original''
Hamiltonians \eqref{eq:HSE_a_FP}
in the polariton basis,
the renormalization problem does not emerge.

As mentioned in the comment,
the calculation based on the Maxwell's boundary conditions
performed in our paper \cite{Bamba2013MBC}
is certainly cumbersome for practical application.
We instead propose the following approach:
Instead of Eq.~\eqref{eq:HSE_standard},
the ``original'' system-environment coupling Hamiltonian should be supposed
as Eq.~\eqref{eq:HSE_H_FP},
which can be derived by the procedure in our recent work \cite{Bamba2013SEC}.
Applying the RWA on Eq.~\eqref{eq:HSE_H_FP}
[or Eqs.~\eqref{eq:HSE_a_FP} when we specify the matters and the gauge]
in the basis of true eigen-states of the cavity system
(polariton modes in the present case) \cite{Beaudoin2011PRA,Bamba2013SEC},
we can derive the master equation \cite{Beaudoin2011PRA},
quantum Langevin equations \cite{Ciuti2006PRA},
and the loss rates can also be derived.
This procedure is applicable to any cavity structures
(whose ``original'' system-environment coupling Hamiltonians
can be derived \cite{Bamba2013SEC}) in principle
and to any matter systems interacting with the electromagnetic fields ultrastrongly,
and we do not need the renormalization required by starting from Eq.~\eqref{eq:HSE_standard}.
Instead of the cumbersome treatment in our paper \cite{Bamba2013MBC},
the complexity is moved to the derivation of the ``original''
system-environment coupling Hamiltonians,
which is in principle determined by the detailed mechanism
of the confinement and loss of the cavity fields \cite{Bamba2013SEC}.
As discussed in our recent work \cite{Bamba2013SEC},
these complexities are reduced to the emergence of another degree of freedom:
whether the system-environment coupling is mediated by the electric field
or the magnetic one,
which are distinguished clearly in the ultrastrong light-matter coupling regime.

This work was supported by KAKENHI 24-632
and the JSPS through its FIRST Program.


\begin{thebibliography}{10}
\expandafter\ifx\csname href\endcsname\relax\def\href#1#2{#2}\fi

\bibitem{DeLiberato2013a2}
S.~De~Liberato, \href{http://arxiv.org/abs/1307.5615}{arXiv:1307.5615
  [quant-ph]}.

\bibitem{Bamba2013MBC}
M.~Bamba and T.~Ogawa,
  \href{http://link.aps.org/doi/10.1103/PhysRevA.88.013814}{Phys. Rev. A
  {\bf 88}, 013814 (2013)}.

\bibitem{Gunter2009N}
G.~Gunter, A.~A. Anappara, J.~Hees, A.~Sell, G.~Biasiol, L.~Sorba,
  S.~De~Liberato, C.~Ciuti, A.~Tredicucci, A.~Leitenstorfer, and R.~Huber,
  \href{http://dx.doi.org/10.1038/nature07838}{Nature {\bf 458}, 178 (2009)}.

\bibitem{Schwartz2011PRL}
T.~Schwartz, J.~A. Hutchison, C.~Genet, and T.~W. Ebbesen,
  \href{http://link.aps.org/doi/10.1103/PhysRevLett.106.196405}{Phys. Rev.
  Lett. {\bf 106}, 196405 (2011)}.

\bibitem{Todorov2009PRL}
Y.~Todorov, A.~M. Andrews, I.~Sagnes, R.~Colombelli, P.~Klang, G.~Strasser, and
  C.~Sirtori,
  \href{http://link.aps.org/doi/10.1103/PhysRevLett.102.186402}{Phys. Rev.
  Lett. {\bf 102}, 186402 (2009)}.

\bibitem{Scalari2012S}
G.~Scalari, C.~Maissen, D.~Tur\v{c}inkov\'a, D.~Hagenm\"uller, S.~De~Liberato,
  C.~Ciuti, C.~Reichl, D.~Schuh, W.~Wegscheider, M.~Beck, and J.~Faist,
  \href{http://www.sciencemag.org/content/335/6074/1323.abstract}{Science {\bf
  335}, 1323 (2012)}.

\bibitem{Johansson2010PRA}
J.~R. Johansson, G.~Johansson, C.~M. Wilson, and F.~Nori,
  \href{http://link.aps.org/doi/10.1103/PhysRevA.82.052509}{Phys. Rev. A {\bf
  82}, 052509 (2010)}.

\bibitem{Bamba2013SEC}
M.~Bamba and T.~Ogawa, \href{http://arxiv.org/abs/1306.2099}{arXiv:1306.2099
  [quant-ph]}.

\bibitem{Beaudoin2011PRA}
F.~Beaudoin, J.~M. Gambetta, and A.~Blais,
  \href{http://link.aps.org/doi/10.1103/PhysRevA.84.043832}{Phys. Rev. A {\bf
  84}, 043832 (2011)}.

\bibitem{Ciuti2006PRA}
C.~Ciuti and I.~Carusotto,
  \href{http://link.aps.org/doi/10.1103/PhysRevA.74.033811}{Phys. Rev. A {\bf
  74}, 033811 (2006)}.

\end{thebibliography}

\end{document}